\documentclass[superscriptaddress,twocolumn,showpacs,showkeys,pra,amsmath,nofootinbib]{revtex4}

\usepackage{graphicx}
\usepackage{xcolor,hyperref}

\newcommand{\ket}[1]{|#1\rangle}
\newcommand{\bra}[1]{\langle#1|}
\newcommand{\braket}[2]{\langle#1|#2\rangle}

\newcommand{\eq}{\begin{equation}}
\newcommand{\fine}{\end{equation}}

\newcommand{\tr}[1]{\text{Tr}[#1]}
\begin{document}


\title{
{Hyperentanglement witness}
}
\author{Giuseppe Vallone}
\homepage{http://quantumoptics.phys.uniroma1.it/}
\affiliation{
Dipartimento di Fisica della Sapienza Universit\`{a} di Roma,
Roma, 00185 Italy}
\affiliation{
Consorzio Nazionale Interuniversitario per le Scienze Fisiche della Materia,
Roma, 00185 Italy}
\author{Raino Ceccarelli}
\homepage{http://quantumoptics.phys.uniroma1.it/}
\affiliation{
Dipartimento di Fisica della Sapienza Universit\`{a} di Roma,
Roma, 00185 Italy}
\author{Francesco De Martini}
\homepage{http://quantumoptics.phys.uniroma1.it/}
\affiliation{
Dipartimento di Fisica della Sapienza Universit\`{a} di Roma,
Roma, 00185 Italy}
\affiliation{
Consorzio Nazionale Interuniversitario per le Scienze Fisiche della Materia,
Roma, 00185 Italy}
\affiliation{Accademia Nazionale dei Lincei, via della Lungara 10, Roma 00165, Italy}
\author{Paolo Mataloni}
\homepage{http://quantumoptics.phys.uniroma1.it/}
\affiliation{
Dipartimento di Fisica della Sapienza Universit\`{a} di Roma,
Roma, 00185 Italy}
\affiliation{
Consorzio Nazionale Interuniversitario per le Scienze Fisiche della Materia,
Roma, 00185 Italy}

\date{\today}

\begin{abstract}
A new criterium to detect the entanglement present in a {\it hyperentangled state}, based on the evaluation of an entanglement witness, is presented. 
We show how some witnesses recently introduced for graph states, measured by only two local settings, can be used in this case.
We also define a new witness $W_3$ that improves the resistance to noise by increasing the number of local measurements.
\end{abstract}

\pacs{03.67.Mn, 03.65.Ud}
\keywords{hyperentanglement, entanglement witness}
\maketitle

\section{Introduction}
Quantum entanglement represents the basic property underlying
many quantum computation processes and quantum cryptographic 
schemes. It guarantees in principle secure cryptographic 
communications and a huge speedup of some important 
computation tasks. In this respect entanglement represents 
the basis of the exponential parallelism of the future quantum 
computers. 

By using optical techniques the entanglement was realized in 
many experiments, either with quantum states based on two \cite{kwia95prl}, 
four \cite{wein01pra,bour04prl,kies05prl}, or even six photons \cite{lu07nap}, 
or, more recently, with multiphoton states, containing more than 
$10000$ entangled particles \cite{dem08prl}.

By hyperentanglement more degrees of freedom (DOF's) of the photons 
are involved and entangled states spanning a high-dimension Hilbert space 
can be created \cite{barb05pra,05-cin-rea,05-bar-gen,04-bar-gen}. 
A hyperentangled (HE) state encoded in $n$ DOF's is expressed by 
the product of $n$ Bell states, one for each DOF.
Double Bell HE states of two photons (i.e. with $n=2$) are currently realized 
in the laboratories and enable to perform tasks that are usually not 
achievable with normally entangled states. Among many applications, 
the realization of a complete Bell state analysis \cite{barb07pra,schu06prl,07-wei-hyp},
and the recently realized enhanced dense coding\cite{barr08nat}, 
are particularly worth of noting. 
By operating with HE states of two photons and $n$ independent 
DOF's we are able to encode the information in $2n$ qubits. 
This significatively reduces the typical decoherence problems 
of multiphoton states based on the same number of qubits and 
dramatically increases the detection efficiency. 
HE states of increasing size are also important for the
realization of advanced quantum nonlocality tests and represent a 
viable resource to increase the power of computation of a 
scalable quantum computer operating in the one-way model \cite{brie01prl, raus01prl}.
Indeed it has been recently demonstrated that efficient $4$-qubit $2$-photon cluster 
states are easily created starting from $2$-photon HE states   
\cite{vall07prl, chen07prl, vall08prl, vall08lpl}.
\begin{figure}[t]
\includegraphics[width=7cm]{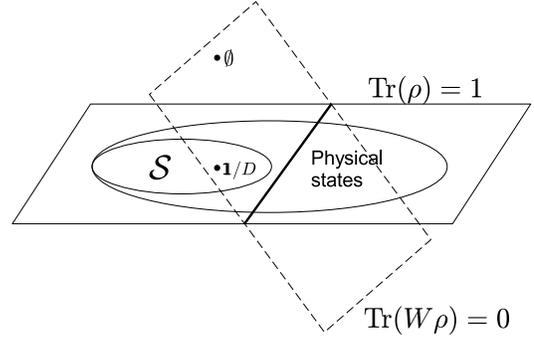}
\caption{Geometrical representation of witnesses in the Hilbert space of hermitian operators $\rho$. 
In this space the scalar product is defined by $\langle\rho,\sigma\rangle\equiv\text{Tr}[\rho^\dag\sigma]$. 
Physical states (i.e. the unit norm positive hermitian operators) 
lie on the hyperplane defined by $\langle \openone,\rho\rangle=\text{Tr}(\rho)=1$. The convex set $\mathcal S$ represents 
the separable states. 
We show also the completely mixed state $\openone/D\in\mathcal S$.
The (hermitian) witness $W$ identifies a hyperplane (passing through the null state $\emptyset$) 
by the equation $\langle W,\rho\rangle\equiv\text{Tr}(W\rho)=0$ and splits the whole Hilbert space in two subsets, one of them
containing $\mathcal S$. All the separable states $\sigma\in\mathcal S$ thus satisfy the condition $\langle\sigma,W\rangle\geq0$ and lie on one side of the hyperplane.
The physical states satisfying the condition $\langle\rho,W\rangle<0$ lie on the other side and thus are entangled.
}
\label{fig:witness}
\end{figure}

The analysis of multiqubit entangled states performed by quantum state 
tomography is particularly demanding since the number of required measurements 
scales exponentially with the number of qubits. 
Furthermore, in practical realizations entanglement is degraded by decoherence and
by any dissipation processes deriving from the unavoidable coupling with the
environment. Being entanglement an expensive resource, its efficient
detection with the minimum number of measurements is a crucial issue 
and new efficient analysis tools are necessary to characterize 
the entanglement of a particular multipartite state.  
The method of ''entanglement witness'' \cite{horo96pla,00-ter-bel} (see Fig. \ref{fig:witness} 
for a geometrical representation of an entanglement witness), first 
demonstrated for entangled states of two photons \cite{03-bar-det} allows us to 
assess the presence of entanglement by using only a few local
measurements. After its introduction the use of entanglement witness was extended to the detection of entanglement 
of various kinds of four qubit entangled states 
\cite{bour04prl,kies05prl,05-wal-exp,vall07prl} and $N$-qubit cluster states. 
At the same time a big effort was spent in the study of the entanglement 
witness operator properties\cite{00-lew-opt,01-lew-cha,02-bru-ref,02-guh-det,06-bre-opt}
and their non-linear generalization\cite{06-guh-non,08-guh-ite}.
 
By the present paper we address the timely question of finding a criterium 
to determine the presence of entanglement in HE states and introduce an entanglement
witness which enables to detect entanglement in these states.
This criterium is different from those used in case of multiparticle entangled states, where each qubits is encoded
in a different particle. The difference resides in the different partitions that can be made in the two cases as it will be shown in the following section.

\section{Witness for hyperentangled state}
\begin{figure}[t]
\includegraphics[width=4cm]{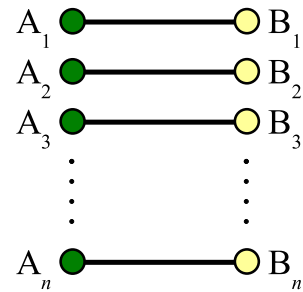}
\caption{(Color online) A Hyperentagled state represented as a graph state where a dots corresponds to a qubit
and the links represent the entanglement existing between the corresponding qubits.
Dark (green) and bright (yellow) dots represent the degrees of freedom of
particles $A$ and $B$ respectively.}
\label{fig:hyper}
\end{figure}

Let's consider the generic DOF $A_j$ ($B_j$), with $j=1,...n$, 
of particle $A$ ($B$).
Each DOF spans a $2$-dimensional Hilbert space (i.e. it is equivalent to a qubit) 
whose basis is $\{{\ket0}_{A_j},{\ket1}_{A_j}\}$ ($\{{\ket0}_{B_j},{\ket1}_{B_j}\}$) for particle $A$ ($B$). 
In this way each particle encodes exactly $n$ qubits.
Let's define
\eq
\mathcal U\equiv\{A_1,B_1,\cdots,A_n,B_n\}
\fine
the set containing the entire number of DOF's.
The (pure) HE state is written as
\eq\label{HE}
\ket\Xi={\ket{\phi^+}}_{A_1B_1}{\ket{\phi^+}}_{A_2B_2}\dots{\ket{\phi^+}}_{A_nB_n}\,,
\fine
where
\eq
\ket{\phi^+}_{A_jB_j}\equiv\frac{1}{\sqrt2}(\ket{0}_{A_j}\ket{0}_{B_j}+\ket{1}_{A_j}\ket{1}_{B_j})
\fine
represents a maximally entangled Bell state.
In general $\ket{\phi^+}$ can be replaced with any maximally entangled state 
(this corresponds to apply single qubit unitaries on the HE state).
In the language of graph states, a HE state can be interpreted as a graph state up to 
a Hadamard gate applied to each qubit $A_j$ (see Fig. \ref{fig:hyper}).
Let's define now the following $N=2n$ operators
\eq
\begin{cases}
 \begin{aligned}
S_{2j}&=Z^{A}_jZ^{B}_j\\
S_{2j-1}&=X^{A}_jX^{B}_j
 \end{aligned}
 \qquad\qquad
j=1,\cdots,n\end{cases}
\fine
where $Z^A_j$, $X^A_j$ ($Z^B_j$, $X^B_j$) are respectively the Pauli matrices $\sigma_z$ and $\sigma_x$ 
acting on the $j$th DOF of particle $A$ ($B$).
By using the stabilizer formalism \cite{gott96pra} the HE state 
can be also defined as the state satisfying
\eq
S_k\ket{\Xi}=\ket{\Xi},\qquad \forall k=1,\cdots, N 
\fine
In general we can define the {\it stabilizer} basis as
\eq
\begin{aligned}
S_k\ket{{\bf s}}=(-1)^{s_k}\ket{{\bf s}}\,\quad\forall k=1,\cdots, N \\
\ket{{\bf s}}\equiv\ket{[s_1,s_2,\cdots,s_N]}\qquad s_k=0,1
\end{aligned}
\fine
and $\ket{\Xi}\equiv\ket{[0,0,\cdots,0]}$.

How can we detect entanglement in this case? And also: 
which kind of entanglement we would like to detect?
The entanglement witness method is based on the introduction of a witness $W$, 
i.e. a hermitian operator whose expectation value is non negative for a 
generic separable state, while it is negative for the entangled state 
we want to detect.
Since the witness is defined up to a multiplicative positive constant, 
here and in the following we fix the normalization 
of a generic witness $W$ for the HE state by requiring that
\eq\label{normalization}
\bra{\Xi}W\ket{\Xi}=-1
\fine
The advantages of this choice will become evident when the resistance to noise
of the entanglement witness will be evaluated (see Section \ref{sec:noise}).

Let's define the entanglement we want to discriminate.
A state $\ket\varphi$ is separable (in the hyperentangled sense) if it satisfies the 
following condition:
\eq\label{partition}
\exists j\text{ such that }\ket{\varphi}=\ket{\varphi_1}_{A_j\mathcal I}\ket{\varphi_2}_{B_j\mathcal J}
\fine
In this equation $\{\mathcal I,\mathcal J\}$ represents a generic bi-partition of the set ${\mathcal T}_j\equiv\{A_1,B_1,\cdots,A_n,B_n\}\backslash\{A_j,B_j\}$, 
so that $\mathcal I\cup\mathcal J=\mathcal T_j$ and $\mathcal I\cap\mathcal J=\emptyset$.

{{\bf Definition:} A (mixed) state is defined to be {\it hyperentangled} in $n$ degrees of freedom if 
it is separately entangled in each of them and cannot be written as a mixture of states that satisfy \eqref{partition}.}

In this way the possibility that a classical mixture of two or more 
states that are not entangled in each DOF, such as those satisfying eq. \eqref{partition},
can be interpreted as hyperentangled, is avoided. 
This definition of separability is different from the usual definition used in the multiparticle entanglement, 
where separability is referred to every possible partition of the qubits. In the hyperentanglement case a state is separable
if it can be expressed by a partition that separates the same DOF, as written in equation \eqref{partition}.
This condition is weaker then the usual multiparticle entanglement as it can be immediately seen by looking at the state $\ket{\Xi}$ in equation \eqref{HE}:
this is a hyperentangled state if $A_1,\cdots A_n$ ($B_1,\cdots B_n$) refer the the different DOF's of particle $A$ ($B$).
However if any $A_j$ (or $B_j$) represents a different particle, the state $\ket{\Xi}$ is no more completely entangled 
(or ``genuine multiparticle entangled'').

{ The first condition is the entanglement in every DOF. We can then measure $n$ entanglement witnesses. 
For each DOF $j$ we can introduce a witness given by}\footnote{
The choice 
of this witness at this stage is arbitrary. Other witnesses, such as $\mathcal W^{(j)}=\openone_2-2{\ket{\phi^+}}_{A_jB_j}{\bra{\phi^+}}=
\frac12(\openone_2-S_{2j}-S_{2j-1}-S_{2j}S_{2j-1})$ can be chosen.
The witness chosen in equation \eqref{W_j} is useful in terms of the measurement settings, as shown in the following.}: 
\eq\label{W_j}
{\mathcal W^{(j)}=\openone_2-S_{2j}-S_{2j-1}.}
\fine
If all of them are negative,
\eq\label{multi-ent}
\tr{\mathcal W^{(j)}\rho}<0\qquad \forall j=1,\cdots,n
\fine 
we know that there is entanglement in each DOF.
Hence all the reduced matrices 
\eq
\rho_j=\text{Tr}_{\mathcal U\backslash {A_jB_j}}[\ket\Xi\bra\Xi]\,,
\fine
obtained by tracing all the DOF's but $A_j$ and $B_j$,
are entangled.

In case of pure states the previous condition assures that each DOF of $A$ 
is entangled with the corresponding one of $B$.
However  this condition, {although necessary,} is not sufficient to demonstrate that the state 
cannot be created by a classical (i.e. mixed) superposition of states that are not 
entangled in all the DOFs.
This is clearly explained by a simple example.
Let's consider the case with $n=2$ and two states
\eq\label{psi12}
\begin{aligned}
\ket{\psi_1}&={\ket{0}}_{A_1}{\ket{0}}_{B_1}{\ket{\phi^+}}_{A_2B_2}\\
\ket{\psi_2}&={\ket{\phi^+}}_{A_1B_1}{\ket{0}}_{A_2}{\ket{0}}_{B_2}
\end{aligned}
\fine
{They are not entangled in every DOF since $\ket{\psi_1}$ ($\ket{\psi_2}$) is separable in the first (second) DOF.
Then the two states are not hyperentangled:}
\eq\label{vev}
\begin{aligned}
&\bra{\psi_1}\mathcal W^{(2)}\ket{\psi_1}=\bra{\psi_2}\mathcal W^{(1)}\ket{\psi_2}=-1\,,\\
&\bra{\psi_1}\mathcal W^{(1)}\ket{\psi_1}=\bra{\psi_2}\mathcal W^{(2)}\ket{\psi_2}=0\,.
\end{aligned}
\fine
However by taking the mixture of these two states with equal weights, 
$\rho'=\frac12\ket{\psi_1}\bra{\psi_1}+\frac12\ket{\psi_2}\bra{\psi_2}$, we obtain
\eq\label{rho'}
\tr{\mathcal W^{(1)}\rho'}=\tr{\mathcal W^{(2)}\rho'}=-\frac12
\fine
In this case the mixture of two non HE states { is entangled in every DOF.}
Note that this feature doesn't depend on the particular choice of the witness 
$\mathcal W^{(j)}$ in eq. \eqref{multi-ent}.

{The correct identification of a hyperentangled state can be obtained} by introducing {a {\it hyperentanglement witness} 
for the HE state $\ket{\Xi}$:}
\eq\label{widetildeW}
\begin{aligned}
\widetilde W&=\openone-2\ket\Xi\bra\Xi\\
&=\openone-\frac{2}{2^N}\sum_{\{s_k\}}S^{s_1}_1S^{s_2}_2\cdots S^{s_N}_N
\end{aligned}
\fine
In the Appendix we will show that 
\eq
|\braket{\Xi}{\varphi}|^2\leq\frac12
\fine
for all the states $\ket{\varphi}$ that satisfy eq. \eqref{partition}.
{The expectation value of $\widetilde W$ is thus negative for $\ket\Xi$ but is positive for all the states expressed in the form \eqref{partition}, 
and then it is positive for all their mixtures. For example, it can be easily verified 
that for the state $\rho'$ in \eqref{rho'} it holds
$\tr{\widetilde W\rho'}=0$ and hence $\rho'$ is (correctly) not hyperentangled.}

Given a witness $\widetilde W$, other witnesses $W'$ can be derived on the basis of the following argument \cite{05-tot-det, toth05pra}:
if we can find a constant $\alpha>0$ such that the operator 
$\mathcal O\equiv W'-\alpha\widetilde W$ is
positive definte (i.e. $\text{Tr}[\mathcal O\rho]\geq0$, $\forall\rho$), 
then $W'$ is a witness.
In fact if $\text{Tr}[W'\rho]<0$ on a generic state $\rho$ 
it comes out that $\text{Tr}[\widetilde W\rho]\leq\frac1\alpha\text{Tr}[W'\rho]<0$ 
and thus $\rho$ is entangled. 

Following this observation, it is shown in \cite{05-tot-det, toth05pra} 
that the following operators
\begin{align}
W_1=(N-1)-\sum^N_{k=1}S_k\,,\qquad\qquad\\
W_2=3-2\left(\prod_{\text{odd }k}\frac{S_k+1}{2}+\prod_{\text{even }k}\frac{S_k+1}{2}\right)
\end{align}
are witnesses for a generic connected graph state.
The same argument can be repeated here, since it uses only 
the stabilizer equation.
In fact, even in the case of a HE state (or whatever state defined in terms of the  stabilized equation)
the operators $W_1-\widetilde W$ and 
$W_2-\widetilde W$ are positive definte.
This is simply checked in the {\it stabilizer} basis $\ket{{\bf s}}$ 
where they are diagonal and thier eigenvalues are non-negative.

We can also introduce here an other witness by using the same argument:
\eq\label{W3}
W_3=2-3\prod^n_{j=1}\left(\frac{1+S_{2j-1}+S_{2j}}{3}\right)
\fine
In order to demonstrate that $W_3$ is a witness, let's consider $W'_3=c_0-3\prod^n_{j=1}\left(\frac{1+S_{2j-1}+S_{2j}}{3}\right)$ and
calculate the lowest eigenvalues of $\mathcal O_3\equiv W'_3-\alpha \widetilde W$. 
If they are positives then $\mathcal O_3$ is positive definite.
The lowest eigenvalues are $\lambda_1=c_0-\alpha-3+2\alpha$ for $\ket{\Xi}$ and $\lambda_2=c_0-\alpha-1$ 
for a state with only one $s_k$ equal to  $1$.
They are equal when $\alpha=1$ and are both positive when $c_0\geq2$.

The witness \eqref{W3} is built by considering all the 
possible products of stabilizers where, for each DOF, we can measure $\openone$, $XX$ or $ZZ$.

The four witnesses of above differ each other with respect to the number of measurement settings, 
i.e. their {\it local decompositions} \cite{toth05pra}
are different.
We remember that the local decomposition of $W$ is defined by the equation $W=\sum_i\mathcal M_i$, where
$\mathcal M_i$ is measured by a different local measurement setting 
$\{O^{(k)}\}^N_{k=1}$. Each $\mathcal M_i$ then consists of the simultaneous measurements 
of $O^{(k)}$ on the corresponding qubit $k$.

In the case of $W_1$ and $W_2$ the local decomposition consists of two terms: 
the first is computed by the local setting
{$X_1X_2X_3X_4...$ and the second by $Z_1Z_2Z_3Z_4...$}.
Hence $W_1$ and $W_2$ require, as shown in \cite{toth05pra} only two local settings,
while the number of settings needed for $\widetilde W$ and $W_3$ scales exponentially with $n$
(at most we need $3^n$ setting for $\widetilde W$ and $2^n$ measurement settings for $W_3$) \cite{05-tot-det}.

{
It is worth noting that the witness introduced in \eqref{W_j} can be measured by the same local settings 
needed for the measurements of $W_1$, $W_2$ (and of course by those used for $W_3$ and $\widetilde W$).
}

Moreover the advantage given by a low number of measurement settings for the witness $W_1$ and $W_2$
is paid by their weakness with respect to the resistance to noise. This will be shown in the following section.

\section{Resistance to noise}\label{sec:noise}
The strength of a witness if often measured by its resistance to noise, 
i.e. the amount of noise that can be added to
the entangled state in such a way that the witness still measures it as entangled.
Consider the following states
\eq
\rho=(1-p_\text{noise})\ket{\Xi}\bra{\Xi}+p_\text{noise}\frac{1}{D}\openone
\fine
where we have defined $D=2^N$ and $p_\text{noise}$ measures the amount of (white) noise present in the state.
The expectation value of a generic witness (normalized such that $\bra{\Xi}W\ket{\Xi}=-1$) is given by
\eq
\text{Tr}[W\rho]=-1+p_\text{noise}+p_\text{noise}\frac{1}{D}\text{Tr}[W]
\fine
so the state is entangled if
\eq
\quad p_\text{noise}<\frac{D}{\text{Tr}[W]+D}\equiv p_\text{M}
\fine
where $p_M$ is the maximum allowed amount of noise.
The trace of $W$ is thus a good measurement of the weakness of the witness: the lower the trace the stronger the resistance to noise.
In table \ref{table:pM} we show the traces of the above defined 
witnesses in order of the increasing resistance to noise.
While $\widetilde W$ is highly resistent to white noise ($p_M$ is always greater than $50\%$) but requires many measurement settings, $W_1$ and $W_2$
requires only two measurement settings but they are less resistent to noise.
For any value of $N$ the resistance to noise of our defined witness \eqref{W3} 
is larger than the witness $W_2$ and $W_1$, and $W_3$ can be seen as a compromise between the need of 
lowering the number of settings and that of increasing the noise resistance.
In general (for any $N$) the noise tolerance of $W_3$ is at least $33\%$.
In conclusion, the resistance to noise of the witness grows with the number of
settings needed to evaluate it.

\begin{table}
\begin{ruledtabular}
\begin{tabular}{ccc}
Witness & Tr[$W$] & $p_M$\\
$W_1$ &$(N-1)D$ & $\frac1{N}=\frac1{\log_2D}\geq0$\\
$W_2$(even $N$) & $3D-4\sqrt D$ & $\frac14\frac{1}{1-\frac1{\sqrt D}}\geq\frac14$\\
$W_2$(odd $N$) & $3D-3\sqrt{2 D}$ & $\frac14\frac{1}{1-\frac3{4\sqrt {2D}}}\geq\frac14$\\
$W_3$(even $N$) & $2D-\frac{3D}{(\sqrt D)^{\log_23}}$ & $\frac13\frac{1}{1-\frac{1}{(\sqrt D)^{\log_23}}}\geq\frac13$\\
$W_3$(odd $N$) & $2D-\frac{3D}{2(\sqrt {D/2})^{\log_23}}$ & $\frac13\frac{1}{1-\frac{1}{2(\sqrt {D/2})^{\log_23}}}\geq\frac13$\\
$\widetilde W$ & $D-2$  &  $\frac12\frac{1}{1-\frac 1D}\geq\frac12$
\end{tabular}
\end{ruledtabular}
\caption{Traces of the witness operator defined in the text and their corresponding resistance to the white noise. 
The number $p_M$, depending on the dimension $D$ of the Hilbert space, represents the maximum amount of noise tolerated by the 
{corresponding hyperentanglement witness}.}
\label{table:pM}
\end{table}

\section{New entanglement witness for generic graph states}
The same witness defined in \eqref{W3} can be used 
for a generic graph state $\ket{G_N}$ associated to a connected graph of $N$ vertices.
The graph state is defined by the stabilizer equation as
\eq
S_k\ket{G_N}=\ket{G_N},\qquad \forall k=1,\cdots, N 
\fine
where
\eq
S_k=X_{k}\prod_{j\in \mathcal N_k}Z_{j}\qquad\quad k=1,\cdots,N\\
\fine
Here $X_{k}$ ($Z_{k}$) are the Pauli matrix $\sigma_x$ ($\sigma_z$) acting on qubit $k$
and $\mathcal N_k$ is the set of qubits to which it is linked.
As shown in \cite{bour04prl}, for a given connected graph, the 
following witness detects the genuine N-qubit entanglement of the corresponding
graph state $\ket{G_N}$\footnote{This witness is slightly different from that defined in 
\cite{bour04prl}, due to the normalization constant needed to satisfy \eqref{normalization}.}:
\eq
\widetilde W=\openone-2\ket{C_N}\bra{C_N}
\fine
Each state of the form $\ket{\phi_1}_{\mathcal A}\ket{\phi_2}_{\mathcal B}$ (where $\mathcal A,\mathcal B$ is a generic partition of the $N$ qubits)
has a positive expectation value of this witness.

Following the same arguments of the previous section 
it is possible to show that the followign operator is a witness

\eq
W_3=
\begin{cases}
2-3{\displaystyle\prod^n_{j=1}}\left(\frac{1+S_{2j-1}+S_{2j}}{3}\right)\qquad N=2n\\
 2-3{\displaystyle\prod^n_{j=1}}\left(\frac{1+S_{2j-1}+S_{2j}}{3}\right)\frac{1+S_{N}}{2}\qquad N=2n+1
\end{cases}
\fine
The number of settings needed for $\widetilde W$ scales exponentially with the number of qubits \cite{05-tot-det},
while we need more than two measurement settings to determine $W_3$.
The maximum amount of noise $p_M$ allowed for $W_3$ is shown in table \ref{table:pM}.

\section{conclusions}

In {this paper we have introduced a method to detect if a 2-particle state is hyperentangled.
The method is based on the initial detection of entanglement for each separate degree of freedom.
Once this condition is satisfied the negative value of a {\it hyperentanglement witness}
operator detects the hyperentanglement.
We introduced four different hyperentanglement witnesses, namely $\widetilde W$, $W_1$, $W_2$ and $W_3$. 
They are characterized by 
different values of the resistance to noise which grows with the number of measurement settings needed 
for their evaluation.
Precisely, only two local settings are required in the case of $W_1$ and $W_2$, while for $W_3$ and $\widetilde W$ their number 
scales exponentially with the number of DOFs. It is worth noting that the local setting used to measure $\widetilde W$, $W_1$, $W_2$ and $W_3$ can be also used to
measure the witnesses $\mathcal W^{(j)}$ (see equation \eqref{W_j}) and then reveal the entanglement corresponding to each DOF, separately.
For example with the local setting $Z_1Z_2Z_3Z_4Z_5Z_6\dots$ it is possible to measure the observables $S_2$, $S_4$, $S_6$, $\dots$ and all their products.

A low number of measurements corresponds to a reduced resistance to noise. Indeed, it is shown that the amount of white noise tolerated by
$W_3$ and $\widetilde W$ exceeds those tolerated by $W_1$ and $W_2$ (see table \eqref{table:pM}).
}

In general, a hyperentangled state \eqref{HE} can also be expressed as a maximally 
entangled state of two qu$d$it, where $d=2^n$, since each particle encodes $n$ qubits. 
However our approach is different from the usual bipartite qudit entanglement 
since the witness detecting the bipartite entanglement between the two particles
is (up to normalization) $W=\frac{\openone}{2^n}-\ket\Xi\bra\Xi$ since the maximum overlapp between a maximally 
entangled state of two qu$d$it and a separable state is exactly $\frac1{2^n}$. 
The witness \eqref{widetildeW} is far more stringent since many entangled states 
in the bipartite sense are not hyperentangled. For example the state $\ket{\psi_1}$ in \eqref{psi12}
can be written as $\ket{00}_A\ket{00}_B+\ket{01}_A\ket{01}_B$. This state clearly shows entanglement between $A$ and $B$ but it is not
hyperentangled.

\begin{acknowledgments}
This work was supported by Finanziamento di Ateneo 06, Sapienza Universit\`a di Roma.   
\end{acknowledgments}

\appendix
\section{Demonstration}
Let the state $\ket{\varphi}$ satisfy eq. \eqref{partition}. This means that $j$ must exist such 
that $\ket\varphi$ can be written as
\eq
\begin{aligned}
\ket{\varphi}=&\ket{\varphi_1}_{A_j\mathcal I}\ket{\varphi_2}_{B_j\mathcal J}\\
=&(a\ket0_{A_j}\ket{\chi_1}_\mathcal{I}+b\ket1_{A_j}\ket{\chi_2}_\mathcal{I})\\
&\otimes(c\ket0_{B_j}\ket{\chi_3}_\mathcal{J}+d\ket1_{B_j}\ket{\chi_4}_\mathcal{J})
\end{aligned}
\fine
with $\ket{\chi_i}$ normalized and $|a|^2+|b|^2=|c|^2+|d|^2=1$. Let's define 
$\gamma_{\mu\nu}$ the overlapp between $\ket{\chi_\mu}_\mathcal{I}\ket{\chi_\nu}_\mathcal{J}$
and 
${\bra{\phi^+}}_{A_1B_1}\dots{\bra{\phi^+}}_{A_{j-1}B_{j-1}}{\bra{\phi^+}}_{A_{j+1}B_{j+1}}\dots{\bra{\phi^+}}_{A_{n}B_{n}}$.
Since the states $\ket{\chi_i}$ are normalized we have $|\gamma_{\mu\nu}|^2\leq1$.
By calculating the overlapp we obtain
\begin{widetext}
\eq
\begin{aligned}
|\braket{\Xi}{\varphi}|^2&=\bra{\phi^+}\left(ac\gamma_{13}\ket{00}+ad\gamma_{14}\ket{01}+bc\gamma_{23}\ket{10}+bd\gamma_{24}\ket{11}\right)_{A_jB_j}
=\frac12|(ac\gamma_{13}+bd\gamma_{24})|^2
\\
&\leq\frac12\left(|(ac\gamma_{13}|^2+|bd\gamma_{24}|^2\right)\leq\frac12\left(|ac|^2+|bd|^2\right)\leq\frac12\left(|a|^2+|b|^2\right)=\frac12
\end{aligned}
\fine
where we used $|\gamma_{\mu\nu}|^2,|c|^2,|d|^2\leq1$ and the property $|r+s|^2\leq|r|^2+|s|^2$.
The bound is easily saturated, for example by the states of the form
\eq
{\ket{\phi^+}}_{A_1B_1}\dots{\ket{\phi^+}}_{A_{j-1}B_{j-1}}{\ket{0}}_{A_{j}}{\ket0}_{B_{j}}{\ket{\phi^+}}_{A_{j+1}B_{j+1}}\dots{\ket{\phi^+}}_{A_{n}B_{n}}.
\fine
\end{widetext}


\end{document}